\documentstyle[12pt]{article}

\begin{document}
\begin{titlepage}
\thispagestyle{empty}
\title{Evolution, probabilistic interpretation and decoupling  
of orbital and total angular momenta in nucleon}
\vspace{4.0cm}
\author{O.V. Teryaev \\
\small \em Bogoliubov Laboratory of Theoretical Physics, \\
\small \em Joint Institute for Nuclear Research,\\
\small \em 141980 Dubna, Moscow region,  Russia\footnote{Permanent address} \\
\small \em and\\
\small \em Centre de Physique Th\'eorique, 
{\footnote {unit\'e propre 014 du
CNRS}}\\ 
\small \em Ecole Polytechnique \\
\small \em F91128 Palaiseau, France}

\date{March  1998}

\maketitle

\begin{abstract}
The leading order evolution of parton orbital angular momenta is treated  
in the probabilistic manner. 
As a result, the splitting probability matrix, while coinciding
with the recent explicit calculations,
is expressed in the terms of the standard 
spin-dependent and spin-averaged kernels. 
The implied relations between spin and orbital 
angular momenta evolution kernels may be interpreted as resulting from the 
ambiguity in the separation of the angular momenta to the
spin and orbital contributions and are likely to be valid for higher 
orders and twists. 
Consequently, the orbital and total angular momenta may be considered 
as decoupling from the spin ones, the latter being the only 
elements of nucleon structure accessible via the hard
scattering of polarized particles. 
In particular, the partition of the total angular momenta of 
quarks and gluons should be analogous to that of 
their momentum at any $Q^2$,
generalizing the earlier finding for asymptotic limit 
by Ji, Tang and Hoodbhoy.
\end{abstract}
%The resulting restrictions 
%for the higher orders and higher twist contributions to evolution
%are discussed.    

 %\noindent Key-Words : Parton distributions, polarized inequality
%constraints

%\smallskip

%\noindent Number of figures : 3

\smallskip

%\noindent March 1998

%\noindent CPT-97/P.3538

%\noindent Web address: www.cpt.univ-mrs.fr
\end{titlepage}

The orbital angular momenta of partons are known to be the
necessary counterparts of their spins QCD evolution \cite{Ratcl}. 
In particular, they are 
responsible for the cancellation of the large gluon helicity 
generated during the evolution, as was confirmed by the
explicit calculation \cite{Ji}, generalized recently to the higher
moments of the orbital angular momenta distributions \cite{HS,HK}. 
At the same time, it 
remains unclear, which general properties of the anomalous dimensions,
if any, are relevant for such a cancellation. 

The more general question is the ambiguity of the splitting 
of the total angular momentum to the spin and orbital parts \cite
{Bel,JM,OT92}. In particular, the choice of energy momentum tensor
suggested by Belinfante \cite{Bel} allows to present the total angular
momentum like the orbital one. 
One may ask, how such an ambiguity is compatible with the QCD 
evolution.  
     
In the present paper, the orbital angular momenta evolution 
is rederived using the kinetic interpretation \cite{CQ} of the 
evolution equations \cite{GL,AP,D} which appeared to be 
especially useful for the analysis of the spin dependence 
\cite{BST,BLT}. As a result, the orbital angular momentum evolution
is expressed entirely in the terms of the spin-dependent and
spin-averaged kernels. While the conservation of total angular
momentum is automatic, 
the extra relation between spin-dependent and spin-averaged kernels 
\begin{eqnarray}
\int_0^1 dx x \Delta P_{Gq}(x)=
{1\over 2} \int_0^1 dx x P_{Gq}(x),
\label{c}\end{eqnarray}
%results in the following relation between the splitting kernels matrices
guarantees, that the evolution of total angular momenta is equal to that of
orbital angular momenta. This proves the compatibility of the 
QCD evolution with the Belinfante procedure. We discuss the related
constraints for the NLO and power corrections to 
the angular momenta evolution. 
The emerging picture is a sort of decoupling of the 
spin and orbital/total angular momenta. While the first manifest
themselves in the hard spin-dependent processes, the latter are the
by-products of the momentum distribution and therefore related to
unpolarized scattering.
  In particular, the similarity between evolution of parton momenta and
angular momenta should result in the similarity of their partition at
any $Q^2$,

The key element of the kinetic approach is the treatment of the virtual gluon
contribution as a flow of the partons {\it from} the given point, while the 
terms associated to real gluons  are interpreted as a flow {\it to} that
point. As a result, the conservation of the operator like vector
current or energy-momentum tensor acquire a form of particle number 
conservation. 

Say, the evolution equations for the singlet case,
after writing \cite{CQ} the virtual corrections to the splitting
kernels like  
\begin{eqnarray}
P_+ (z) =P(z)-\delta(1-z)\int_0^1 P(t) dt,
\label{+}\end{eqnarray}
and performing the change of variable $t=y/x$
in the last integral
(which allows to interpret the partons decrease from virtual corrections 
as flow to the points $y<x$, similarly to that from real corrections),
may be written as a Master equation of the Boltzmann type 
%(see, e.g. \cite{VK})
\begin{eqnarray}
{d [x f_i (x)] \over{dt}}=
\sum_{j} \int_0^1 dy ( w_{ij} (y \to x) y f_j (y)-
w_{ji} (x \to y) xf_i (x)).
\label{sk}\end{eqnarray}

Here $i,j$ are labels for $q,G$, and the transition rates per unit
time $t=ln Q^2$ 
\begin{eqnarray}
 w_{ij}(y \to x) = {\alpha_s \over {2 \pi}}
 P_{ij} ({x \over y})
  {{x \theta (y > x)} \over {y^2}}.
\label{sigmasp}\end{eqnarray}

The probabilistic interpretation of the orbital angular momentum
evolution appeared already in the pioneering papers \cite{Ratcl,Ji}.
Say, the splitting of the quark to quark and gluon should be
accompanied by the production of the orbital angular momentum,
balancing the helicity of gluon (as the quark helicity is conserved).
To calculate the net effect, one should subtract the probabilities 
of having the gluon with negative and positive helicities. 
Note that just the same combination (up to a sign)
appears when one is calculating the 
gluon-quark spin-dependent splitting kernel, with the momentum fraction
$1-x$:
 
\begin{eqnarray}
P^{LS}_{qq}(x)+P^{LS}_{Gq}(1-x) = - \Delta P_{Gq} (1-x),
\label{lqqg}\end{eqnarray}
where the upper inices are showing that the orbital
angular momentum is produced by the spin one. 
The expression in the l.h.s appeared because the produced orbital
angular momentum is carried by the quark with a momentum fraction
$x$ and gluon with a momentum fraction $1-x$. 
It is possible to find the ratio of the quark and gluon orbital
angular momenta by means of the entirely probabilistic reasoning as well. 
Namely, suppose that the quark momentum before splitting has only 
$z$ component, while the momenta of the final partons are in the 
$x-z$ plane. The $z$ components of the orbital angular momenta are 

  \begin{eqnarray}
L_z^q= P_x r^q_y;\ \L_z^G= -P_x r^G_y,
\label{lz}\end{eqnarray}
where the $x$ components of quarks and gluon momenta are equal,
up to a sign, due to the momentum conservation, and the effective 
spatial non-locality of quark and gluon production $r^{q,G}$
is introduced. The latter is leading also to the appearance of the 
$x$ components of the orbital angular momenta: 
\begin{eqnarray}
L_x^q= -P^q_z r^q_y;\ \L_x^G= -P^G_z r^G_y.
\label{lx}\end{eqnarray}

Due to the conservation of the $x$ component of the angular momentum 
$L_x^q= -L_x^G$, and, consequently;
\begin{eqnarray}
{{r^q_y} \over {r^G_y}}=-{{P_z^G} \over {P_z^q}}.
\label{rl}\end{eqnarray}
Note that the contribution of the $x$ components of the parton
helicities to the angular momentum conservation is resulting in 
the appearance of the additional terms of the order of $P_x^2/P_z^2$
in the r.h.s. of (\ref{rl}) which should be neglected at leading order.
Substituting (\ref{rl}) to  (\ref{lz}) one get the partition rule
\begin{eqnarray}
{{L_z^q} \over {L_z^G}}={{P_z^G} \over {P_z^q}}={{1-x} \over x},
\label{pr}\end{eqnarray}
coinciding with the one found \cite{Ji} by the explicit calculation.

Although our derivation is a classical one and included the
effective non-locality $r$, the latter did not enter the final
result, and so due to the correspondence principle the above
derivation, based on the conservation laws, is applicable in the 
actual quantum case as well. 
This would result in the substitution of the nonlocality $r$ by the 
derivative operator in the momentum space. The simultaneous use of the
analogs of (\ref{lz}, \ref{lx}) would mean the similarity in the
action of such a derivative to the (non-forward) matrix
elements of the $T^{0x}$ and $T^{0z}$ components of energy-momentum
tensor, required by the Lorentz invariance. 

Applying the partition rule (\ref{pr}), one may deduce from (\ref{lqqg})
\begin{eqnarray}
P^{LS}_{qq}(x) = (x-1) \Delta P_{Gq} (1-x), \nonumber \\
P^{LS}_{Gq}(x) = (x-1) \Delta P_{Gq} (x).
\label{lqg}\end{eqnarray}
%where we changed in the last equation $x \to 1-x$. 
Let us pass to the contribution to the orbital angular momentum 
due to the gluon
helicity. The case of the quark-antiquark pair
production is the 
most simple one, as their helicities are opposite in the chiral limit
and all the gluon helicity is converted to the (anti)quark orbital momentum:
\begin{eqnarray}
P^{LS}_{qG}(x) = (1-x) P_{qG} (x),
\label{lgq}\end{eqnarray}
where the $1-x$ factor is coming form the partition ratio(\ref{pr}),
which does not depend on the nature of the splitted and produced partons.

Consider now the case of gluon-gluon splitting, when
the helicities of all the participating gluons should be fixed:
\begin{eqnarray}
P^{LS}_{GG}(x) = (1-x) (P^{+,+-}_{GG}(x)+P^{+,-+}_{GG}(x)-P_{GG}^{+,++}(x)),
\label{lgg}\end{eqnarray}
where the partition factor is appearing again, 
and the helicity of the
gluon with the momentum fraction $x$ is listed second, while the
incoming gluon helicity is first, as usual. Actually, all the
quantities in the r.h.s. were already calculated in the pioneering
paper \cite{AP} as an intermediate step of the computation of the
spin-dependent kernel. However, it is instructive to use instead the
general expressions:
\begin{eqnarray}
P_{GG}(x) = P^{+,+-}_{GG} (x)+P^{+,-+}_{GG}(x)+P_{GG}^{+,++}(x), \nonumber \\
\Delta P_{GG}(x) = P^{+,+-}_{GG} (x)-P^{+,-+}_{GG}(x)+P_{GG}^{+,++}(x)
\label{asgg}\end{eqnarray}
and the symmetry conditions
\begin{eqnarray}
P^{+,+-}_{GG} (x)=P^{+,-+}_{GG}(1-x), \
P_{GG}^{+,++}(x)=P_{GG}^{+,++}(1-x).
\label{sygg}\end{eqnarray}
Combining these equations one easily get 
\begin{eqnarray}
P^{LS}_{GG}(x) = (x-1) (\Delta P_{GG} (x)+\Delta P_{GG} (1-x)-P_{GG} (x)).
\label{lggf}\end{eqnarray}  

Finally, the splitting matrix has the form:
\begin{eqnarray}
&&P^{LS}(x)=-\bar x \cdot  \left(
\begin{array}{cc}
 \Delta P_{Gq}(\bar x) & - 2 n_f P_{qG} (x) \\
 \Delta P_{Gq}(x)& \Delta P_{GG} (x)+\Delta P_{GG} (\bar x)-P_{GG} (x)
\end{array}
\right),
\label{matr}\end{eqnarray}
where $\bar x = 1-x$ and 
the obvious dependence on the flavour number is
restored.

Note that all the kernels in (\ref{matr}) do not have a singular
parts. It is a common feature 
in the kinetic approach, where the latter are
treated separately via (\ref{+}). It is natural 
therefore that  (\ref{matr})
do not contain a singular terms at all. Recall, that such a terms 
are emerging due to the
infrared divergence and are appearing in the diagonal kernels only. 
The resulting expression exactly coincides with the recent explicit
calculation \cite{HS,HK}. However, 
the properties of the orbital
angular momentum evolution can be seen now in a more general way. 

Namely, the evolution of the sum of the quark and gluon  angular
momenta, generated by the quark and gluon helicities, has a form:
\begin{eqnarray}
\frac{d}{dt} (L_q+L_G)=\int_0^1 dx (x-1) (\Delta P_{Gq}(1-x) + 
\Delta P_{Gq}(x)) \cdot \int_0^1 dx \Delta \Sigma (x) + \nonumber \\
\int_0^1 dx (x-1) (\Delta P_{GG} (x)+\Delta P_{GG} (1-x)-P_{GG} (x)-
2 n_f P_{qG}(x))
\cdot \int_0^1 dx \Delta G (x).
\label{cons}\end{eqnarray}
Changing $x \to 1-x$ in the second term of the quark coefficient and 
substituting $1-x \to 1/2$ for the gluonic coefficient, which is
possible because of the symmetry of the integrated function with 
respect to the interchange of $x$ and $1-x$, one get:
\begin{eqnarray}
\frac{d}{dt} (L_q+L_G)=-\int_0^1 dx \Delta P_{Gq}(x)
\cdot \int_0^1 dx \Delta \Sigma (x) - \nonumber \\
\int_0^1 dx (\Delta P_{GG} (x)-x P_{GG} (x)-2 n_f x P_{qG}(x))
\cdot \int_0^1 dx \Delta G (x).
\label{cons1}\end{eqnarray}
The coefficients of $\Delta \Sigma$ and $\Delta G$ coincide, up to a
sign, with those appearing in the evolution of gluon helicity
(while quark helicity is conserved), guaranteeing the conservation of
the total angular momentum. Note that the terms in the orbital angular
momentum evolution, proportional to the unpolarized splitting kernels,
are canceled by the terms in the gluon helicity evolution, coming
from the virtual corrections, which are the same for the
spin-dependent and spin-independent case (\ref{sk}).     

To proceed further, one should note that the diagonal evolution of the
orbital angular momenta, generated by themselves, also does not include
any new ingredients. As it was shown \cite{Ji} for the first moment 
and generalized recently 
by H\"agler and Sch\"afer \cite{HS} for the higher moments, this 
coincides with the evolution of unpolarized densities, with the extra
power of $x$,
$P^{LL}_{ij}(x)=x P_{ij}(x)$,
so that the diagonal evolution of orbital angular momenta
is precisely the same as that of the energy momentum tensors:
$P^{LL(1)}_{ij}=P^{TT(2)}_{ij}$, 
where the moment is labeled in brackets.
%This fact is also natural in the probabilistic interpretation.

Combining all these facts, 
another interesting property of evolution may be found. 
Namely, the evolution of the {\it total}
angular momenta $J_q={1 \over 2} \Delta \Sigma +L_q, J_G=\Delta G+
L_G$ acquires the simple form:
\begin{eqnarray}
&&\frac{d}{dt}
\left(
\begin{array}{ll}
 J_q\\
 J_G \\
\end{array}
\right)
= \frac{\alpha(t)}{2\pi} 
\left(
\begin{array}{cc}
 \int_0^1 dx (x-1) P_{qq}(x)  &  2 n_f \int_0^1 dx x P_{qG}(x) \\
 \int_0^1 dx x  P_{Gq}(x)  &  -2 n_f \int_0^1 dx x P_{qG}(x)
\end{array}
\right)
\left(
\begin{array}{ll}
 J_q\\
 J_G\\
\end{array}
\right)
\label{j}\end{eqnarray}
and is governed by the same matrix $P^{LL(1)}=P^{TT(2)}$. 
This result was first mentioned in \cite{Ji1,Ji2} as a result of 
explicit calculation \cite{Ji} and was adjusted to the proportionality
of the relevant operators. 

One should note that the equality 
\begin{eqnarray}
P^{JJ(1)}_{ij}=P^{LL(1)}_{ij} 
\label{Bel}\end{eqnarray}
may be also considered as a
quantitative expression of the Belinfante invariance, allowing 
for the representation of the total angular momentum in the orbital form.  
To understand this fact further, let us study, which properties of the
kernels are implied by (\ref{Bel}). 
It is easy to find the following relation between the matrices 
for the spin and orbital angular momenta evolution:
 \begin{eqnarray}
\tilde P_{ij}^{LS(1)}+\tilde
P_{ij}^{SS(1)}=P_{ij}^{LL(1)}(=P_{ij}^{JJ(1)}=
P_{ij}^{TT(2)}), \nonumber \\
 P_{ij}^{SL(1)}=0.
\label{constr}\end{eqnarray}
Here $\tilde P$ mean the evolution kernels, where distributions
are weighted with their spin factors, which practically is 
resulting in the substitution $\tilde P^{SS}_{qG}(x)=1/2 P^{SS}_{qG}(x), 
\tilde P^{SS}_{Gq}(x)=2 P^{SS}_{Gq}(x), \tilde P^{LS}_{qq}=2 P^{LS}_{qq},
\tilde P^{LS}_{Gq}(x)=2 P^{LS}_{Gq}(x)$, all other entries 
of $\tilde P(x)$ and $P(x)$ being equal. 
Note that these general relations
%,being the main result of the present paper, 
imply the conservation of the total angular momentum as well, 
as it is now the consequence of the momentum conservation. 

%One may check that (\ref{constr}) are actually valid at leading order,
%(which can be already seen from the results of \cite{Ji}, where 
%the similar evolution of $T$ and $J$ is also mentioned and discussed.
%although relations of the type (\ref{constr}) 
%are not written down explicitely).
The general reasons for such an equalities are the symmetry properties of
the kernels with respect to the interchange of $x$ and $1-x$, 
and the extra relation, required for the validity of the 
$qq$ and $Gq$ components of (\ref{constr}).
\begin{eqnarray}
 \int_0^1 dx x \Delta P_{Gq}(x)={1 \over 2} \int_0^1 dx x P_{Gq}(x),
\label{constr1}\end{eqnarray}
where $1/2$ in the r.h.s. is just the quark spin appeared when 
transformation to $\tilde P$ is performed.
This equality  may be easily checked explicitly. One may however ask,
are there
any general reasons for it. 
Such an explanation would be especially desirable, because of the
relation to the Belinfante invariance, mentioned above. 

As soon as the origin of (\ref{constr}) is rather general, 
these relations should be satisfied for the higher orders of 
perturbation theory (where the whole probabilistic interpretation
and the symmetry properties are, generally speaking, lost), 
as well as for the higher 
twists\footnote{In particular, the two-loop contribution of axial anomaly 
to $P^{SS}_{qq}$ is cancelled by the relevant term in $P^{LS}_{qq}$,
which may be identified by calculation of the gluonic matrix elements 
of quark orbital momentum \cite{OT90,Ji}.}.

Of course,  
this hypothesis should be checked by further explicit calculations.
In particular, higher loop calculations of $P_{SS}$ and $P_{LS}$
within dimensional regularization  
contain the extra ambiguity due to the choice of $\gamma_5$ matrix,
and so it may either cancel identically in their sum, or imply some 
particular factorization scheme, preserving Belinfante invariance,
like it happen for Adler-Bardeen \cite{Cheng,MT} 
relation or conformal invariance\cite{M}. 

In what follows, we shall suppose the validity 
of (\ref{constr}) and 
outline its possible physical consequences. 

The partition of the total angular momentum should 
{\it always} follow the partition of momentum: 
\begin{eqnarray}
{{J_q(Q^2)} \over{J_G(Q^2)}}=
{{\sum_f \int_0^1 dx x q_f (x,Q^2)} \over {\int_0^1 dx x G (x,Q^2)}}.
\label{constr2}\end{eqnarray}
As a result, at
moderate $Q^2$ gluons should carry about half of the total angular
momentum, as was anticipated by X. Ji \cite{Ji2}, 
starting from the similar evolution and the 
results of sum rule calculation \cite{BJ}. 
The Belinfante invariance is the new argument in favor of such a 
situation and one might look 
for the general non-perturbative proof of equality (\ref{constr2}), based on
the related arguments. 

Such a behavior would mean, that spin and orbital/total angular momenta 
are less dependent, when one might expect from the very beginning. 
The orbital and total angular momenta 
are completely defined by the momentum distribution,
manifested in the scattering of the {\it unpolarized} particles and
are in this
sense decoupled from the spin ones. Recall, that total angular momentum 
conservation, being the main link between spin and orbital angular
momenta, is actually guaranteed by momentum conservation and 
Belinfante invariance (\ref{constr}).
At the same time, the role of spin angular momenta in their canonical
form is coming from
their relation to OPE and manifestation in {\it spin-dependent} processes.
 
In such a case, one should not expect any new information about 
nucleon spin structure from Deeply Virtual Compton Scattering 
\cite{Ji1} \footnote{Although this process was originally suggested 
for this purpose, its subsequent studies \cite{DVCS}, 
establishing a new rapidly developing branch of PQCD, 
lost this original connection.}. 
Instead, one may consider the relation of $J$ and $T$ as an extra
constraint for the relevant non-forward parton distribution,
namely: 
\begin{eqnarray}
 \sum_f \int_0^1 dx x E(x,\Delta)=0,
\label{vcs}\end{eqnarray}
which is just the term potentially making a difference between momentum and 
total angular momentum partition \cite{Ji1}.
As the first moment of this quantity
\begin{eqnarray}
 \int_0^1 dx E(x,\Delta)=F_2(0)
\label{vcs1}\end{eqnarray}
describes the anomalous magnetic 
moment of nucleon, the 
smallness of its isoscalar part is making such a
behavior 
rather natural. 
Moreover, Belinfante invariance is providing the natural explanation of that
fact (c.f. \cite{OT90,OT92})!
At the same time,
model approaches \cite{Seh} may provide the estimate of the effective quark
orbital angular momentum as well as
the higher moments of the total angular momenta,
where (\ref{constr}) are not valid, and which are a natural 
objects for the models applications. Note also, that 
orbital momentum in field theory may essentially differ from that in
quantum mechanics and in the potential models.  

Let us also briefly discuss the implications of another constraint in 
(\ref{constr}), $P_{SL}=0$. 
This would mean that the orbital angular momentum is never
produced by the spin one. At the same time, the possible gluon
fusion at low $x$ \cite{GLR}
should give rise to such a production. 
One should conclude, that either this effect is oscillating, 
changing sign to give 
a zero in (\ref{constr}) after taking the first moment, 
or absent at all for some reasons.  
Note also, that this constraint is more general in a sense that it
requires the similar evolution
of $L$ and $J$, while both may differ from that of $T$.

In conclusion, we have found a representation for the leading order 
orbital angular momentum 
evolution, where its probabilistic nature is fully clear.
The relevant splitting kernels are expressed via the usual spin-averaged
and spin-dependent ones and are suitable for the numerical calculations.  
The implied constraints for  the kernels are related to the ambiguity of 
separation of orbital and spin degrees of freedom and are expected
to be valid for the higher order and power corrections. 
The resulting picture of the nucleon spin structure is rather simple.
The orbital and (averaged over $x$) total 
momenta are decoupled and completely determined
by the unpolarized scattering. The polarized scattering is entirely
related to spin angular momenta, which should be studied in details  
in the forthcoming polarization experiments; the latter would be a
final judgment for the proposed picture.  
The theoretical checks 
of the latter should include the study of orbital 
angular momentum within NLO approximations, higher twists corrections
and non-perturbative approaches. 

\vspace*{7mm}
I am indebted to A.V. Efremov, E. Leader, B. Pire, P.G. Ratcliffe,
A. Sch\"afer and J. Soffer for stimulating discussions. 
This research is partly performed in the framework of the Grants
96-02-17631 of Russian Foundation for Fundamental Research
and Grant $N^o_-$ 93-1180 from INTAS.

\end{document}